\documentclass[amsmath,amssymb,aps, showpacs,
twocolumn
]{revtex4-1}

\usepackage{graphicx}%
\usepackage{bm}%

\begin{document}

\title{Tunneling-assisted optical information storage with lattice polariton solitons in cavity-QED arrays}

\author{E.~S.~Sedov$^{1}$, A.~P.~Alodjants$^{1,2}$, S.~M.~Arakelian$^{1}$}
\affiliation{$^{1}$Department of Physics and Applied Mathematics, Vladimir State University named after A. G. and N. G. Stoletovs, Vladimir, 600000, Russia\\
$^{2}$Russian Quantum Center, 100~ Novaya str., Skolkovo, Moscow region, 143025, Russia}

\author{You-Lin Chuang$^{3}$, YuanYao Lin$^{3}$, Wen-Xing Yang$^{3,4}$, and Ray-Kuang Lee$^{3}$}
\affiliation{$^{3}$Institute of Photonics Technologies, National Tsing-Hua University, Hsinchu, 300, Taiwan\\
$^{4}$Department of Physics, Southeast University, Nanjing 210096, China}

\begin{abstract}
Considering  two-level media in the array of weakly coupled nano-cavities, we reveal a variety of dynamical regimes, such as diffusion, self-trapping, soliton, and breathers for the wave-packets in the presence of photon tunneling processes between the next-nearest cavities.
We focus our attention on the low branch (LB) bright polariton soliton formation,  due to the two-body polariton-polariton scattering processes.
When detuning frequency is manipulated  adiabatically,  the low-branch lattice polariton localized states, {\it i.e.}, that are  solitons and breathers evolving between photon-like and matter-like states, are shown to act as carriers for  spatially distributed storage and retrieval of optical information.
\end{abstract}


\maketitle

\section{\label{intro}INTRODUCTION}
Nowadays, the elaboration and investigation of hybrid quantum devices and artificial  nanostructures represent a huge area of experimental and theoretical research~\cite{Zheludev2010,SimonAfzeliusAppelEurPhysJD}.
In particular, quantum memory devices are proposed for mapping the quantum state of light onto the matter by using a slow light phenomenon, through the coupling between matter excitation and quantized field~\cite{SimonAfzeliusAppelEurPhysJD,HammererSorensenPolzik2010}.
In this sense, polaritons, linear superpositions of quantized field and collective excitations in matter, provide a very elegant way for optical information storage, where the group velocity of the wave-packet could be low enough due to a large value of polariton mass~\cite{FleischhauerLukin2002,Leksin2007}.

Within the framework of modern scalable quantum technologies~\cite{Jaksch,JiangBrennen2008,Daley,TomadinFazio2010},  the arrays of cavities containing two-level systems (atoms, quantum dots, or Cooper pair boxes, referred as qubits) strongly interacting with a cavity field at each site, are theoretically supposed to provide a promising platform for quantum computing  and quantum information processing~\cite{Hartmann,LeiLee2008,AngelakisSantos2007}.
Moreover, a strong Kerr-type nonlinearity caused by two-body polariton-polariton interaction leads to the formation of bright polariton solitons~\cite{AmoPigeonSanvitto2011,SichKrizhanovskiiSkolnick2012,ChenLinLai2012}.

In the experiments, great efforts have been aimed at the achievement of deterministic trapping of  single qubits with a strong  coupling to the  quantized  electromagnetic  fields in  nanocavities~\cite{FaraonMajumdarVuckovicPRL1042010,SchlosserReymondProtsenkoNature4112001, VetschReitzSaguePRL1042010, ThompsonTieckeScience3402013}.
Especially, we stress here the recent challenging results established  in Ref. ~\cite{ThompsonTieckeScience3402013} with  ``ultracold''  single rubidium  atoms trapped in the vicinity of  tapered fiber (about 100 nm far from) and their effective coupling with photonic crystal cavity.
The obtained single photon Rabi frequency was in the range of few gigahertz for the cavity volume less than $\lambda ^{3}$  ($\lambda$ is light wavelength).
Such results pave the way to the design of new scalable devices for quantum memory purposes being compatible with photonic circuits~\cite{ChutinanandSajeevOptExp2006}; these devices  exploring two-level systems at their heart~\cite{Leksin2007,HeshamiGreenYangHanPRA862012}.

Here, we apply full power of current quantum technological achievements obtained in the atomic optics area  to provide theoretically an alternative approach to optical information storage and retrieval by using half-matter, half-photon property of polaritons and by investigating collective dynamics of coupled atom-light states in a qubit-cavity quantum electrodynamical (QED) array.
Low branch (LB) polariton solitons, as well as different dynamical regimes for diffusion, self-trapping, and breather states occur through the interaction between atoms and quantized optical cavity field~\cite{TrombettoniSmerziPRL2001,WangZhangZhangMaXuePRA810336072010}.
Considering the next-nearest tunneling effect  for photonic fields while the distance between adjacent cavities is within the order of optical wavelength,  lattice polariton soliton solutions are revealed to exist at the border of  two different kinds of breather states.
Due to the robustness in preserving the shape of wave-packets,  by manipulating the detuning frequency adiabatically, optical information storage and retrieval are proposed to carry out through the transformation between photon-like and matter-like lattice polariton solitons.

This paper is arranged as follows. In Sec.~\ref{Model}, we explain in details our model to realize atom-light interaction in a cavity array occurring at nanoscales.
The extended tight-binding model will be established in this case, and 
In Sec.~\ref{POLARITONSINTHENANOSCALE}, we introduce coupled atom-light excitation basis that is polariton basis for  lattice  system.
Apart from results obtained by us previously~\cite{ChenLinLai2012}, LB polaritons occurring at each site of the cavity array are a subject of our study at the rest of the paper.
In Sec.~\ref{TIMEDEPENDENTVARIATIONALAPPROACH}, we use time dependent variational approach to obtain  polariton wave-packet behavior.
Basic equations for the wave-packet parameters and their general properties in the QED cavity array are established.
In Sec.~\ref{polWPdyn}, we establish results relaying  to investigation of variety of 1D lattice polariton wave-packet dynamical regimes in the presence of next neighbor interaction in the lattice.
In Sec.~\ref{Quantumstorageofopticalinformation}, we propose the physical algorithm of storage of optical information by using lattice polariton localized states  that is soliton and breather states.
In the conclusion, we summarize the results obtained.

\section{\label{Model}Model of atom-light interaction beyond the tight-binding approximation}
\begin{figure}
\centering
\includegraphics[width=8.4cm]{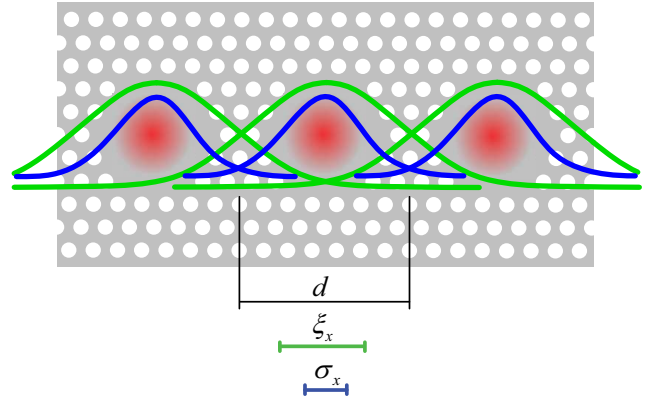}
\caption{\label{FIG_schematicdrawing} (Color online) Schematic for our proposed 1D cavity- QED array, in which each cavity contains the ensemble of 2-level atoms as qubits. Cavities are formed by the defects in a photonic waveguide crystal, with period $d$ that, in fact, represents  characteristic size of cavity; $\xi _{x}$ and $\sigma _{x}$ are   characteristic spatial scales of the optical field and atomic wave function localization, respectively. In this work,  we assume that $\sigma _{x} < \xi _{x} \le d$.}
\end{figure}

We consider a one-dimensional (1D) array of small (nanoscale) cavities, each containing the ensemble of a small but macroscopic number $N_{n}$ of \emph{non-interacting} two-level atoms, see~Fig.~\ref{FIG_schematicdrawing}.
The proposed structure in Fig.~\ref{FIG_schematicdrawing} can be realized by loading a small number of ultracold atoms via a dipole trap, slightly above the so-called collisional blockade regime which is practically valid for the beam waist $w_{0} \ge 1 \mu$m~\cite{ThompsonTieckeScience3402013,SchlosserReymondGrangierPRL2002}.
The total Hamiltonian $\hat{H}$ for our atom-light coupled system in Fig.~\ref{FIG_schematicdrawing} can be represented as~\cite{TomadinFazio2010,BarinovJPB2010,ChenLinLai2012}

\begin{equation}
\label{MainHam}
\hat{H} = \hat{H}_{\rm AT} +\hat{H}_{\rm PH} +\hat{H}_{\rm I},
\end{equation}
where $\hat{H}_{\rm AT} $ is a quantum field theory Hamiltonian for non-interacting atoms, $\hat{H}_{\rm PH} $ is responsible for the photonic field distribution, and the term $\hat{H}_{\rm  I} $ characterizes the atom-light interaction in each cavity, respectively. In the second quantized form, the Hamiltonian in Eq. (1) can be written as
\begin{subequations}
\label{Ham_TSL_PH_I}
\begin{eqnarray}
&&\hat{H}_{\rm AT} = \sum _{ ^{i,j=1,2}_{i\ne j} }\int \hat{\Phi} _{j}^{\dag } \left(-\frac{\hbar ^{2} \Delta }{2M_{at} } +V_{\rm ext}^{(j)}  \right) \hat{\Phi} _{j} d\vec{r}, \label{Ham_TSL}
\\
&&\hat{H}_{\rm PH} = \int \hat{\Phi} _{\rm ph}^{\dag } \left(-\frac{\hbar ^{2} \Delta }{2M_{\rm ph} } +V_{\rm ph} \right) \hat{\Phi} _{\rm ph} d\vec{r},  \label{Ham_PH}
\\
&&\hat{H}_{\rm I} = \hbar \kappa \int \left(\hat{\Phi} _{\rm ph}^{\dag } \hat{\Phi} _{1}^{\dag } \hat{\Phi} _{2} + \hat{\Phi} _{2}^{\dag } \hat{\Phi} _{1} \hat{\Phi} _{\rm ph} \right) d\vec{r}, \label{Ham_I}
\end{eqnarray}
\end{subequations}
with the mass of free atoms,  $M_{\rm at} $, and the effective mass of trapped photons,  $M_{\rm ph} $. In Eq. (\ref{Ham_TSL_PH_I}),  quantum field operators $\hat{\Phi} _{1,2}^{} (\vec{r})$ ($\hat{\Phi} _{\rm ph}$) and  $\hat{\Phi} _{1,2}^{\dag } (\vec{r})$ ( $\hat{\Phi} _{\rm ph}^{\dag }$) annihilate and create atoms (photons) at position $\vec{r}$; $V_{\rm ext}^{(j)} \, (j=1,2)$ and $V_{\rm ph} $ are trapping potentials for atoms and photons, respectively. Since each cavity contains a small number of atoms,  one can safely neglect the terms responsible for atom-atom scattering processes in Eq. (\ref{Ham_TSL})~\cite{AnglinVardiPhysRevA2001}.

In general, one can expand atomic ($\hat{\Phi} _{j}$) and photonic ($\hat{\Phi} _{\rm ph} $) field operators as follows
\begin{subequations}
\label{Phi_at_phot}
\begin{eqnarray}
&&\hat{\Phi} _{j} (\vec{r})=\sum _{n} \hat{a}_{j,n} \varphi _{j,n} \left(\vec{r}\right), \, j=1,2, \label{Phi_at}
\\
&&\hat{\Phi} _{\rm ph} (\vec{r})=\sum _{n} \hat{\psi} _{n} \xi _{n} \left(\vec{r}\right),  \label{Phi_phot}
\end{eqnarray}
\end{subequations}
with the real (Wannier) functions:  $\varphi _{j,n} \left(\vec{r}\right)$, $\xi _{n } \left(\vec{r}\, \right)$, responsible for the spatial distribution of atoms and photons at $n$--site.
They fulfill the normalization condition $\int _{-\infty }^{+\infty }\left(\varphi _{j,n} \left(\vec{r}\right)\right)^{2} d\vec{r}= \int _{-\infty }^{+\infty }\left(\xi _{n} \left(\vec{r}\right)\right)^{2} d\vec{r}= 1$.
Annihilation operators $\hat{a}_{1,n} $ and $\hat{a}_{2,n} $ in Eq. \eqref{Phi_at} characterize the dynamical properties of atoms at internal lower (${\left| 1 \right\rangle} $) and upper (${\left| 2 \right\rangle} $) levels, respectively; meanwhile the annihilation operator $\hat{\psi} _{n} $ in Eq. \eqref{Phi_phot} describing the temporal behavior of the photonic mode located at the $n$-th lattice cavity.

Substituting Eq. (\ref{Phi_at_phot}) for Eq. (\ref{Ham_TSL_PH_I}), one can obtain
\begin{subequations}
\label{Ham_TLS_1_2_12_PH_new_I_new}
\begin{eqnarray}
&&\hat{H}_{\rm AT} = \hat{H}_{1} + \hat{H}_{2} + \hat{H}_{12}, \label{Ham_TLS_1_2_12}
\\
&&\hat{H}_{j} =\hbar \sum _{n}^{M}\left[\omega _{n}^{(j)} \hat{a}_{j,n}^{\dag } \hat{a}_{j,n} -\beta _{j,n} \left(\hat{a}_{j,n}^{\dag } \hat{a} _{j,n+1} + \hat{a}_{j,n}^{\dag } \hat{a} _{j,n-1} \right)   \right], \nonumber \\
&& \hphantom{\hat{H}_{j} =} j=1,2, \label{Ham_1_2}
\\
&&\hat{H}_{\rm PH} =\hbar \sum _{n}^{M}\left[\omega _{n,\, ph} \hat{\psi} _{n}^{\dag } \hat{\psi} _{n} -\alpha _{n}^{(1)} \left(\hat{\psi} _{n}^{\dag } \hat{\psi} _{n+1} + \hat{\psi} _{n}^{\dag } \hat{\psi} _{n-1} \right) \right.
\nonumber \\
&&\phantom{H_{\rm PH} =}
\left. -\alpha _{n}^{(2)} \left(\hat{\psi} _{n}^{\dag } \hat{\psi} _{n+2} + \hat{\psi} _{n}^{\dag } \hat{\psi} _{n-2} \right)\right], \label{Ham_PH_new}
\\
&&\hat{H}_{\rm I} =\hbar \sum _{n}^{M}\frac{g}{\sqrt{N_{n} } } \left[\hat{\psi} _{n}^{\dag } \hat{a}_{1,n}^{\dag } \hat{a}_{2,n} + \hat{a}_{2,n}^{\dag } \hat{a}_{1,n} \hat{\psi} _{n} \right],  \label{Ham_I_new}
\end{eqnarray}
\end{subequations}
where $N_{n} = \hat{a}_{1,n}^{\dag } \hat{a}_{1,n} + \hat{a}_{2,n}^{\dag } \hat{a}_{2,n} $ corresponds to the total number of atoms at the $n$-th lattice cell.
The frequencies $\omega _{n}^{(j)} $, $\omega _{n,\, {\rm ph}} $, hopping constants $\beta _{j,n} $, $\alpha _{n}^{(1)} $, $\alpha _{n}^{(2)} $ and nonlinearity strength $g$ are in the form
\begin{subequations}
\label{beta_alpha12_u_j}
\begin{eqnarray}
&&\omega _{n} ^{(j)} = \frac{1}{\hbar } \int \left(\frac{\hbar ^{2} }{2M_{\rm at} } (\vec{\nabla }\varphi _{j,n})^2   +\varphi _{j,n} V_{\rm ext}^{(j)} \varphi _{j,n} \right)d\vec{r}, \qquad \label{omega_j}
\\
&&\omega _{n, {\rm ph}} = \frac{1}{\hbar } \int \left(\frac{\hbar ^{2} }{2M_{\rm at} } (\vec{\nabla }\xi _{n})^2   +\xi _{n} V_{\rm ph} \xi _{n} \right)d\vec{r}, \qquad \label{omega_phh}
\\
&&\beta _{j,n} = -\frac{1}{\hbar } \int \left(\frac{\hbar ^{2} }{2M_{\rm at} } \vec{\nabla }\varphi _{j,n} \cdot \vec{\nabla }\varphi _{j,n+1} \right.
\nonumber \\
&&\phantom{\beta _{j,n} =}
\left. +\varphi _{j,n}^{} V_{\rm ext}^{(j)} \varphi _{j,n+1} \right)d\vec{r}, \label{beta}
\\
&&\alpha _{n}^{(1)} = -\frac{1}{\hbar } \int \left[\frac{\hbar ^{2} }{2M_{\rm ph} } \vec{\nabla }\xi _{n} \cdot \vec{\nabla }\xi _{n+1} +\xi _{n} V_{\rm ph} \xi _{n+1} \right] d\vec{r},
\nonumber\\ \label{alpha1}
\\
&&\alpha _{n}^{(2)} = -\frac{1}{\hbar } \int \left[\frac{\hbar ^{2} }{2M_{\rm ph} } \vec{\nabla }\xi _{n} \cdot \vec{\nabla }\xi _{n+2} +\xi _{n} V_{\rm ph}^{} \xi _{n+2} \right] d\vec{r},
\nonumber\\ \label{alpha2}
\\
&& g = \kappa \int  \xi _{n} \phi _{1, n} \phi _{2, n}  d\vec{r}. \label{ggggg}
\end{eqnarray}
\end{subequations}

Thereafter, for simplicity we assume that  all cavities are identical to each other and contain the same average number $N=\left\langle N_{n} \right\rangle $ of atoms.
In this case, it is convenient to suppose that functions $\varphi _{j,n} \left(\vec{r}\right)$ are identical for all $n$, that is $\varphi _{j,n} \left(\vec{r}\right)\simeq \varphi _{j,n\pm 1}^{} \left(\vec{r}\right)$.
We are also working under the strong atom-field coupling condition, that is
\begin{equation}
\label{strongcouplingIneq}
g \gg \Gamma _{\rm at}, \Gamma _{\rm ph},
\end{equation}
where $\Gamma _{\rm at}$ and $\Gamma _{\rm ph}$ are spontaneous emission and cavity decay rates, respectively.
The parameter $\alpha _{n}^{(1)} \equiv \alpha ^{(1)} $ in Eq. \eqref{alpha1} describes overlapping of an optical field for the nearest-neighbor cavities; ${\alpha _{n}^{(2)} \equiv \alpha ^{(2)} }$ is responsible for the overlapping of photonic wave functions beyond the tight-binding approximation.
Since the characteristic spatial scale of atomic localization $\sigma _{x}$ is essentially smaller than cavity size $d$, it seems reasonable to use the tight-binding approximation especially for atomic system in the cavity array.
Coupling coefficients $\beta _{j,n} \equiv \beta _{j} $ in Eq. (\ref{beta_alpha12_u_j}) are the nearest-neighbor hopping constants for atoms in a 1D lattice structure.

In particular, we examine the properties of  parameters  for the cavity-QED array containing two-level rubidium atoms.
We take the D-line of rubidium atoms as an example, which gives the resonance frequency ${\omega_{12}\left/ \right. 2\pi = 382 {\rm THz}}$.
The strength of the interaction of a single atom with a quantum optical field is taken as $g_{0} =\sqrt{\frac{\left|d_{12} \right|^{2} \omega _{12} }{2\hbar \varepsilon _{0} V} } \approx 2\pi \times 2.2\, 4{\rm GHz}$ at each cavity with the effective volume of atom-field interaction $V \simeq d^{3} $.
We assume $d=2 {\rm \mu m}$,  that is compatible with current experimental results~\cite{ThompsonTieckeScience3402013}; $d_{12} $ is  the atomic dipole matrix element.
To achieve a strong atom-field coupling regime -- see~Eq.~\eqref{strongcouplingIneq} -- one can propose a macroscopically large number of atoms at each cavity, say $N=100$.
This number  implies a collective atom-field coupling parameter   $g=g_{0} \sqrt{N} \approx 2\pi \times 22.4\, {\rm GHz}$.
The lifetime for rubidium atoms is $27 {\rm ns}$, which corresponds to the spontaneous emission rate $\Gamma _{\rm at}$ of about $2 \pi  \times 6 {\rm MHz}$.
The minimal value of each cavity field decay rate  $\Gamma _{\rm ph}$ can be taken up to several hundred  of megahertz that corresponds to cavity quality factor  $Q\simeq 10^{5} - 10^{6} $.

To get a variational estimate for the tunneling coefficients mentioned above, we assume that  Wannier wave functions for the atomic and photonic parts localized at the \textit{j}th cavity may be approached by
\begin{subequations}
\label{both_phi_xi_gauss}
\begin{eqnarray}
&&\varphi _{j,n} \left(\vec{r}\right)=C_{j} e^{{-\left(x-x_{n} \right)^{2} \mathord{\left/ {\vphantom {-\left(x-x_{n} \right)^{2} 2\sigma _{x,j}^{2} }} \right. \kern-\nulldelimiterspace} 2\sigma _{x,j}^{2} } } e^{{-\left(y^{2} +z^{2} \right)\mathord{\left/ {\vphantom {-\left(y^{2} +z^{2} \right) 2\sigma _{j}^{2} }} \right. \kern-\nulldelimiterspace} 2\sigma _{j}^{2} } },  \label{phi_gauss}
\\
&&\xi _{n} \left(\vec{r}\right)=C_{\xi } e^{{-\left(x-x_{n} \right)^{2} \mathord{\left/ {\vphantom {-\left(x-x_{n} \right)^{2} 2\xi _{x}^{2} }} \right. \kern-\nulldelimiterspace} 2\xi _{x}^{2} } } e^{{-\left(y^{2} +z^{2} \right)\mathord{\left/ {\vphantom {-\left(y^{2} +z^{2} \right) 2\xi ^{2} }} \right. \kern-\nulldelimiterspace} 2\xi ^{2} } }, \label{xi_gauss}
\end{eqnarray}
\end{subequations}
where $C_{j} =\left(\pi ^{3/2} \sigma _{x,j} \sigma _{j}^{2} \right)^{-1/2} $ ($j=1,2$) and $C_{\xi } =\left(\pi ^{3/2} \xi _{x} \xi ^{2} \right)^{-1/2} $ are relevant normalization constants, respectively.
Taking into account the realistic values of atomic and photonic wave functions, we assume
\begin{eqnarray}
\sigma _{x,j} \ll \sigma _{j}, \quad \xi _{x} \ll\xi . \label{sigma_conds}
\end{eqnarray}

If the atoms are trapped in the vicinity of thin (tapered) optical fiber (that is not shown in~Fig.~\ref{FIG_schematicdrawing}), the trapping potential $V_{\rm ext}=V_{\rm w}+V_{\rm opt}$ can be represented as a sum of $V_{\rm w}$ that is Van der Waals potential occuring due to the closeness of  atoms to the fiber surface, and $V_{\rm opt}$ that is a potential created by the optical field,~cf.~\cite{VetschReitzSaguePRL1042010,LacrouteChoiGobanNJP2012}.
For current experiments the depth of total potential $V_{\rm ext}$ is of order of millikelvins~\cite{ThompsonTieckeScience3402013,NotomiKuramochiTanabe}.
Although the general (radial) dependence of $V_{\rm ext}$ on the distance from the surface is not so simple, however, it is possible to consider a harmonic  traping potential at the bottom of  $V_{\rm ext}$ by choosing the appropriate external laser field parameters.
Roughly speaking, we consider atomic trapping potential $V_{\rm ext}$ represented as~ \cite{TrombettoniSmerziSodanoNJP2005}:
\begin{eqnarray}
&&V_{\rm ext } \simeq \frac{M_{\rm at} }{2} \left[ \omega _{x}^{2} \left(x-x_{n} \right)^{2} + \omega ^{2} _{\rm  \bot} \left( y ^{2} + z^{2} \right) \right], \label{pot_opt}
\end{eqnarray}
where $\omega _{ x} $ and $\omega _{\bot }$ are relevant trapping  axial and radial frequencies,  respectively.
We suppose that the minimum of a 2D periodic potential  is  located at a centers $x_{n} =n d$ of the $n$-th cavity.
By  substituting Eq. \eqref{both_phi_xi_gauss} and Eq.~\eqref{pot_opt} into Eq.~(\ref{beta_alpha12_u_j}),  we obtain
\begin{eqnarray}
\label{beta_estimated}
\beta =&&-\frac{\hbar }{4M_{\rm at} \sigma _{x}^{2} } e^{\frac{-d^{2} }{4\sigma _{x}^{2} } } \left(1-\frac{d^{2} }{2\sigma _{x}^{2} } \right),
\end{eqnarray}
for the atomic tunneling rate $\beta$.
The atomic tunneling rate $\beta \equiv \beta _{2}$ is positive if a cavity effective size is ${d > \sqrt{2} \sigma _{x} \approx 1.414\sigma _{x} }$.
The latest one ($\sigma _{x}$) is typically a few hundred nanometers in real experiments with ultracold atoms~\cite{TrombettoniSmerziSodanoNJP2005}.

The calculation of photon tunneling rates $\alpha ^{\left(\zeta \right)} $ (${\zeta =1,2}$) between the cavities can be performed in the same way.
Thus, we have
\begin{equation}
\label{alpha_estimated}
\alpha ^{\left(\zeta \right)} = -\frac{\hbar }{4M_{\rm ph} \xi _{x}^{2} } e^{\frac{-\zeta ^{2} d^{2} }{4\xi _{x}^{2} } } \left(1-\frac{\zeta ^{2} d^{2} }{2\xi _{x}^{2} } \right),
\end{equation}
where $\zeta =1,2$ enumerates the number of cavities.
Taking into account a typical effective photon mass, $M _{\rm ph} \simeq 2.8 \times 10 ^{-36} {\rm kg}$ for rubidium D-lines average wavelength  $\lambda \approx 785{\rm \mu m}$, and assuming the width of photonic wave-packet to be  $\xi = 1 {\rm \mu m}$ for $d=2 {\rm \mu m}$, it is possible to estimate photonic tunneling parameters as $\alpha ^{(1)} \simeq 2 \pi \times 549 {\rm GHz}$ $ (\zeta = 1)$ and $\alpha ^{(2)} \simeq 2 \pi \times 191 {\rm GHz}$ $ (\zeta = 2)$ respectively.

\section{\label{POLARITONSINTHENANOSCALE}Polaritons in the nano-size cavity array}

One of the main features of our approach is a strong nonlinearity due to small cavity volumes occupied by the optical field, that is $V\simeq (\lambda /2n)^{3} $, where $\lambda $ is a  light wavelength, $n$ is a refractive index~\cite{ThompsonTieckeScience3402013,NotomiKuramochiTanabe}.
In Schwinger representation, atom-field interaction in the lattice can be described by atomic excitation operators $\hat{S}_{-,\, n} $, $\hat{S}_{+,\, n} = \hat{S}_{-,n}^{\dag } $ and by operator $\hat{S}_{z,\, n} $ of the population imbalance which are defined as
\begin{eqnarray}
\label{S_minus_plus_z}
&&\hat{S}_{+,\, n} = \hat{a}_{2,n}^{\dag } \hat{a}_{1,n},\\\nonumber
&&\hat{S}_{-,\, n} = \hat{a} _{1,n}^{\dag } \hat{a}_{2,n},\\\nonumber
&&\hat{S}_{z,\, n} =\frac{1}{2} \left( \hat{a}_{2,n}^{\dag } \hat{a}_{2,n} - \hat{a}_{1,n}^{\dag } \hat{a}_{1,n} \right).
\end{eqnarray}
The operators determined in Eq. (12) obey SU(2) algebra commutation relations
\begin{eqnarray}
\left[\hat{S}_{+,\, n}, \hat{S}_{-,\, n} \right] = 2\hat{S}_{z,\, n}, \quad \left[\hat{S}_{z,\, n} , \hat{S}_{\pm, \, n} \right] = \pm \hat{S}_{\pm ,\, n}. \label{komm_rel}
\end{eqnarray}

Alternatively, it is possible to map operators in Eq. (12) onto the atomic excitation operators $\hat{\phi} _{n}$, $\hat{\phi} _{n}^{\dag } $ applying the so-called Holstein--Primakoff transformation, i.e.,
\begin{subequations}
\label{S_plus_new_minus_new_z_new}
\begin{eqnarray}
&&\hat{S}_{+,\, n\, } = \hat{\phi} _{n}^{\dag } \sqrt{N- \hat{\phi} _{n}^{\dag } \hat{\phi} _{n} }, \label{S_plus_new}
\\
&&\hat{S}_{-,\, n} = \sqrt{N- \hat{\phi} _{n}^{\dag } \hat{\phi} _{n} } \hat{\phi} _{n}, \label{S_minus_new}
\\
&&\hat{S}_{z,\, n} = \hat{\phi} _{n}^{\dag } \hat{\phi} _{n} -{N\mathord{\left/ {\vphantom {N 2}} \right. } 2}.  \label{S_z_new}
\end{eqnarray}
\end{subequations}

It is worth noticing that the atomic excitation operators $\hat{\phi} _{n} $ , $\hat{\phi} _{n}^{\dag } $ obey the usual bosonic commutation relations $\left[\hat{\phi} _{\, n} , \hat{\phi} _{m}^{\dag } \right]=\delta _{mn}$.
Strictly speaking, it is possible to treat the operators $\hat{a}_{1,n} $ and $\hat{a}_{2,n}$ describing particles at lower and upper levels, respectively, as $\hat{a}_{1,n} \approx \sqrt{N} -\frac{\hat{\phi} _{n}^{\dag } \hat{\phi} _{n} }{2N^{1/2} } -\frac{\left(\hat{\phi} _{n}^{\dag } \hat{\phi} _{n} \right)^{2} }{8N^{3/2} } $, $ \hat{a}_{2,n} \simeq \hat{\phi} _{n} $~\cite{ChenLinLai2012}.

When  number $N$ at each cavity is macroscopical but not so large,  one can keep all the terms in the expansion of~$\hat{a}_{1,n}$.
In this limit, we get for an effective Hamiltonian $\hat{H} = \hat{H} _{\rm L} + \hat{H} _{\rm TUN} + \hat{H} _{\rm L}$,
\begin{subequations}
\label{Ham_L_C_NL}
\begin{eqnarray}
&&\hat{H}_{\rm L} =\hbar \sum _{n} \left[\tilde{\omega }_{12} \hat{\phi} _{n}^{\dag } \hat{\phi} _{n} +\omega _{n,\, {\rm ph}} \hat{\psi} _{n}^{\dag } \hat{\psi} _{n} +g\left(\hat{\psi} _{n}^{\dag } \hat{\phi} _{n} + H.C.\right)\right],
\nonumber \\ \label{Ham_L}
\\
&&\hat{H}_{\rm TUN} =-\hbar \sum _{n}\left[\beta \left(\hat{\phi} _{n}^{\dag } \hat{\phi} _{n+1} +H.C. \right) \right.
\nonumber\\
&&\phantom{H_{\rm C}}
\left.
+\alpha ^{(1)} \left(\hat{\psi} _{n}^{\dag } \hat{\psi} _{n+1} +H.C. \right)+\alpha ^{(2)} \left(\hat{\psi} _{n}^{\dag } \hat{\psi} _{n+2} +H.C. \right)\right],
\nonumber \\ \label{Ham_C}
\\
&&\hat{H}_{\rm NL} =-\hbar \sum _{n}\left[\frac{g}{2N} \left(\hat{\psi} _{n}^{\dag } \hat{\phi} _{n}^{\dag } \hat{\phi} _{n}^{2}  +H.C. \right)\right], \label{Ham_NL}
\end{eqnarray}
\end{subequations}
where we have introduced new parameters $\tilde{\omega }_{12} =\omega _{n}^{(2)} -\omega _{n}^{(1)} +2\beta _{1,n}$.  Now, let us introduce the lattice polariton operators as follows
\begin{subequations}
\label{Xi_1_Xi_2}
\begin{equation*}
\hat{\Xi} _{1,n} = X_{n} \hat{\psi} _{n} +C_{n} \hat{\phi} _{n},
\,\,\, \hat{\Xi} _{2,n} = X_{n} \hat{\phi} _{n} -C_{n} \hat{\psi} _{n}, \quad \eqno{\rm (\ref{Xi_1_Xi_2}a,b)}
\end{equation*}
\end{subequations}
where $X_{n} $ and $C_{n} $ are Hopfield coefficients defined as
\begin{subequations}
\label{koeff_X_koeff_C}
\begin{eqnarray}
X_{n} =\frac{1}{\sqrt{2} } \left(1+\frac{2 \pi \delta _{n} }{\sqrt{4g^{2} + (2 \pi \delta _{n})^{2} } } \right)^{1/2},\label{koeff_X}
\\
C_{n} =\frac{1}{\sqrt{2} } \left(1-\frac{2 \pi \delta _{n} }{\sqrt{4g^{2} + (2 \pi \delta _{n}) ^{2} } } \right)^{1/2}.\label{koeff_C}
\end{eqnarray}
\end{subequations}

In Eq.~(\ref{koeff_X_koeff_C}), $\delta _{n} =(\omega _{n,\, {\rm ph}} -\tilde{\omega }_{12} )/ 2 \pi $, is atom-light field detuning frequency at each cavity.
Note that we consider parameters $X_{n} $ and $C_{n} $ to be the same for all cavities (sites $n$), assuming that $X\equiv X_{n} $ and $C\equiv C_{n} $.
This approach implies an equal atom-light detuning $\delta =\delta _{n} $ for all cavities too.

The operators $\hat{\Xi} _{1,n} $ and $\hat{\Xi} _{2,n} $ in Eq. (\ref{Xi_1_Xi_2}) characterize two types of bosonic quasiparticles, \textit{i.e.}, upper and lower branch polaritons occurring at each site of the lattice.
At the low density limit, Eqs.~(\ref{Xi_1_Xi_2}-\ref{koeff_X_koeff_C}) represent the exact solution that diagonalizes a linear part $\hat{H}_{\rm L} $ of the total Hamiltonian $\hat{H}$.

\begin{figure}
\centering
\includegraphics[width=8cm]{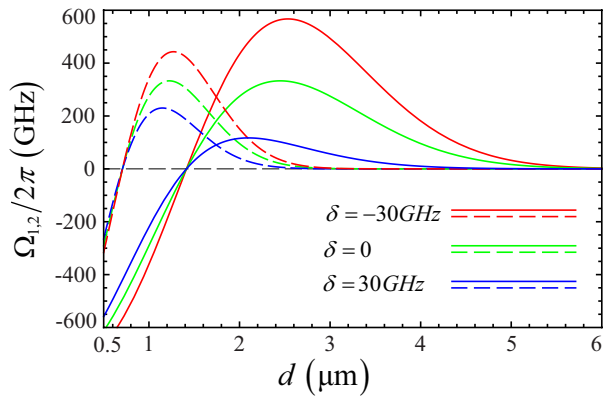}
\caption{\label{FIG_tunn_koeff} (Color online) Tunneling energies between the nearest,  $\Omega _{1} $ (solid curves), and the next-nearest neighbors, $\Omega _{2} $ (dashed curves), are shown as a  function of the size  of  cavity $d$. The widths of wave-functions for cavity field and atoms are estimated as $\xi _{x} =1 {\rm \mu m}$ and $\sigma _{x} =0.4 {\rm \mu m}$, respectively. }
\end{figure}

At equilibrium the lowest polariton branch is much more populated.
Here, we use the mean-field approach to replace the corresponding polariton field operator $\hat{\Xi} _{n} $ by its average value $\left\langle \hat{\Xi} _{n} \right\rangle $, which simply characterizes the LB polariton wave function at the $n$-th cavity.
In particular, for further processing we introduce the $n$-th normalized polariton amplitude  $\Psi _{n} ={\left\langle \hat{\Xi} _{n} \right\rangle \mathord{\left/ {\vphantom {\left\langle \Xi _{n} \right\rangle \sqrt{N_{pol} } }} \right. \kern-\nulldelimiterspace} \sqrt{N_{\rm pol} } } $, where $N_{\rm pol} =\sum _{n}\left\langle \hat{\Xi} _{n}^{\dag } \hat{\Xi} _{n} \right\rangle $ is the total number of LB polaritons at the array.
For this approach,  by substituting Eq. (\ref{Xi_1_Xi_2}) into Eq. (\ref{Ham_L_C_NL}) and keeping LB polariton terms only, we arrive at
\begin{eqnarray}
H &&= \hbar \sum _{n}^{M}\left[
\vphantom{\frac{1}{2}}
\Omega _{\rm LB} \, \left| \Psi _{n} \right|^{2} -\Omega _{1} \left(\Psi _{n}^{*} \Psi _{n+1} +C.C.\right) \right. \nonumber \\
&&\left. -\Omega _{2} \left(\Psi _{n}^{*} \Psi _{n+2} + C.C. \right)  +\frac{1}{2} \Omega _{3} \left|\Psi _{n} \right|^{4} \right], \quad \label{Hamoltonian_norm}
\end{eqnarray}
where we have introduced characteristic frequencies
\begin{subequations}
\label{Omega_1_2_3_4}
\begin{eqnarray}
&&\Omega _{\rm LB} =\frac{1}{2} \left(\tilde{\omega }_{12} +\omega _{n,\,{\rm  ph}} -\sqrt{(2 \pi  \delta) ^{2} +4g^{2} } \right), \label{Omega_LB}
\\
&&\Omega _{1} =\beta X^{2} +\alpha ^{(1)} C^{2}, \label{Omega_1}
\\
&&\Omega _{2} =\alpha ^{(2)} C^{2}, \label{Omega_2}
\\
&&\Omega _{3} =2 g C X^{3} \frac{N_{\rm pol} }{N}. \label{Omega_3}
\end{eqnarray}
\end{subequations}
The nearest and next-nearest tunneling energies, $\Omega_1$ and $\Omega_2$, are shown
in Fig.~\ref{FIG_tunn_koeff}, as a function of the characteristic cavity size $d$ for different detuning frequencies $\delta$.
For a large enough ($d \gg \xi _{x} $) cavity size both tunneling rates $\Omega _{1,2} $ are positive, and condition $\frac{\Omega _{2}}{\Omega _{1}} \simeq \frac{\alpha ^{(2)}}{\alpha ^{(1)}} \approx 4 e^{-3d^{2}/4 \xi ^{2} _{x}} \ll 1 $ is fulfilled, see~Fig.~\ref{FIG_tunn_koeff}.
The overlapping of neighbor polariton wave-functions is a dominant term, and our lattice system is reduced to the typical tight-binding approach.
On the other hand, the properties of polariton tunneling energies change  dramatically for the small sized cavities, $d \approx \xi_x$, where coefficient $\Omega _{2} $ becomes much more important.

The main features of the polaritonic lattice are connected with the properties of atom-light detuning $\delta$.
In the limit of a negative and large atom-light field detuning chosen as $\left|2 \pi \delta \right| \gg g$, $\delta <0$ ${(X\simeq {g\mathord{\left/ {\vphantom {g \left|2 \pi \delta \right|}} \right. \kern-\nulldelimiterspace} \left|2 \pi \delta \right|}}, {C\simeq 1)}$,  LB polaritons behave as photons, \textit{i. e.} ${\Xi _{2,n} \simeq \psi _{n} }$.
Thus, we can represent the parameters (\ref{Omega_1_2_3_4}) as ${ \Omega _{\rm LB} \simeq \omega _{\rm ph} }$, ${ \Omega _{1} =\alpha ^{(1)} }$, ${ \Omega _{2} \approx \alpha ^{(2)} }$, ${ \Omega _{3} =2 N_{\rm pol} {g^{4} \mathord{\left/ {\vphantom {g^{4} N\left|\delta \right|^{3} }} \right. \kern-\nulldelimiterspace} N\left|2 \pi \delta \right|^{3} } }$.
However, in another limit, we can take $\left| 2 \pi \delta \right| \gg g$, $\delta >0$ ($X\simeq 1$, $C\simeq {g\mathord{\left/ {\vphantom {g \delta }} \right. \kern-\nulldelimiterspace} \ 2 \pi \delta } $ ) and LB polaritons behave as excited atoms, \textit{i. e.}, $\Xi _{2,n} \simeq \phi _{n} $.
We readily find the coefficients~(\ref{Omega_1_2_3_4}) as ${ \Omega _{\rm LB} \simeq \tilde{\omega }_{12} }$, ${ \Omega _{1} =\beta +\alpha ^{(1)} g^{2} \mathord{\left/ {\vphantom {g^{4} N\left|\delta \right|^{3} }} \right. \kern-\nulldelimiterspace} (2 \pi \delta) ^{2}  }$, ${ \Omega _{2} =\alpha ^{(2)} g^{2} \mathord{\left/ {\vphantom {g^{4} N\left|\delta \right|^{3} }} \right. \kern-\nulldelimiterspace}(2 \pi \delta )^{2} }$, ${ \Omega _{3} = 2 N_{\rm pol} g^{2} \mathord{\left/ {\vphantom {g^{4} N\left|\delta \right|^{3} }} \right. \kern-\nulldelimiterspace}  2 \pi N \delta }$.

In practice, instead of using Eq. (\ref{Hamoltonian_norm}), it is useful to work with the dimensionless Hamoltonian $H\rightarrow H \hbar \left| \Omega _{1} \right|$ in the form
\begin{eqnarray}
H &&= \hbar \sum _{n}^{M}\left[
\vphantom{\frac{1}{2}}
\omega _{\rm LB} \, \left| \Psi _{n} \right|^{2} -\omega _{1} \left(\Psi _{n}^{*} \Psi _{n+1} +C.C.\right)
\right. \nonumber \\
&&-\left. \omega _{2} \left(\Psi _{n}^{*} \Psi _{n+2} + C.C. \right)  + 2 \sqrt{\pi} \omega _{3} \left|\Psi _{n} \right|^{4} \right], \quad \label{Hamoltonian_dimless}
\end{eqnarray}
where $\omega _{LB} =\Omega _{LB} / \left| \Omega _{1} \right| $, $\omega _{1} =\Omega _{1} / \left| \Omega _{1} \right|  \equiv {\rm sgn}\left( \Omega _{1} \right)$, $\omega _{2} =\Omega _{2} / \left| \Omega _{1} \right| $, $\omega _{3} =\Omega _{3} / 4 \sqrt{\pi}\left| \Omega _{1} \right|  $ are normalized parameters characterizing polariton properties in the cavity QED chain.
Eq. (20)  is the starting model equation for the present work and is used to study the nonlinear phase diagrams, as well as the related optical information storage with lattice polariton solitons in the following sections.

\section{\label{TIMEDEPENDENTVARIATIONALAPPROACH}Time-dependent variational approach}

To analyze different regimes of polaritons in the cavity-QED arrays, we study the dynamical evolution of in-site Gaussian shape wave-packet,
\begin{eqnarray}
\Psi _{n} &&={\mathbb N}\exp \left[-\frac{\left(n - \xi (t)\right)^{2} }{\gamma (t)^{2} } + i p(t)(n- \xi (t)) \right.\nonumber \\
&&+ \left. i \frac{\eta (t)}{2} \left(n - \xi(t)\right)^{2}
\vphantom{\frac{\left(n-\xi(t)\right)^{2} }{\gamma (t)^{2} }}
\right], \label{WaveFunc}
\end{eqnarray}
where $\xi (t)$ and $\gamma (t)$ are the time dependent dimensionless center and  width of the wave-packet, respectively; $p (t)$ is momentum and $\eta \left(t\right)$ is curvature, ${\mathbb N}=\left(\sqrt{2}/\sqrt{\pi } \gamma (t) \right)^{1/2} $ is a normalization constant (a wave-packet amplitude).
Lattice coordinate $x$ relates to the number of sites (cavities) $n$ as $x = n d$.
The wave-packet dynamical evolution can be obtained  from the corresponding Lagrangian density
\begin{eqnarray}
\label{Lagrangian}
L=&&\sum _{n}^{M}\left[\frac{i}{2} \left(\Psi _{n}^{*} \frac{\partial \Psi _{n} }{\partial t} -\Psi _{n} \frac{\partial \Psi _{n}^{*} }{\partial t} \right)-\omega _{LB} \, \left|\Psi _{n} \right|^{2} \right.
\nonumber \\
&&+\omega _{1} \left(\Psi _{n}^{*} \Psi _{n+1} +C.C.\right)+\omega _{2} \left(\Psi _{n}^{*} \Psi _{n+2} + C.C. \right)
\nonumber \\
&&- \left. 2 \sqrt{\pi} \omega _{3} \left|\Psi _{n} \right|^{4} \right].
\end{eqnarray}

By plugging Eq.~\eqref{WaveFunc} intor Eq.~\eqref{Lagrangian}, one can have an effective Lagrangian $\bar{L}$ by averaging the Lagrangian density Eq. \eqref{Lagrangian}, as
\begin{eqnarray}
\label{LagrDensity}
\bar{L} = \left[ p\dot{\xi}-\frac{\dot{\eta} \gamma ^{2} }{8} - \frac{\omega _{3}}{\gamma }
 +\omega _{1} \cos \left( p \right) e^{-\sigma }+ \omega _{2} \cos \left( 2 p \right)e^{-4\sigma } \right], \nonumber\\
\end{eqnarray}
where we made the following denotation $\sigma =\frac{\gamma ^{2} \eta ^{2} }{8} +\frac{1 }{2\gamma ^{2} } $. Dots denote derivatives with respect to dimensionless time $t \rightarrow t/2 \left| \Omega _{1} \right|$. It is remarked  that Eq.~\eqref{LagrDensity} is valid when parameter $ \gamma $ is not too small, that is $\gamma > 1$~\cite{TrombettoniSmerziPRL2001,WangZhangZhangMaXuePRA810336072010}.
With the Lagrangian in Eq. \eqref{LagrDensity}, one can obtain the following variational equations for the canonically conjugate polariton wave-packet parameters
\begin{subequations}
\label{kDot_GammaDot_thetaDot}
\begin{eqnarray}
&&\dot{p} = 0, \label{kDot}
\\
&&\dot{\xi} =  \omega _{1} \sin \left( p \right)e^{-\sigma } + 2\omega _{2} \sin \left( 2 p \right)e^{-4\sigma },\,\,\,\, \label{XDot}
\\
&&\dot{\gamma } = \frac{\gamma \eta }{m^{*} }, \label{GammaDot}
\\
&&\dot{\eta } = \frac{1 }{m^{*} } \left(\frac{4}{\gamma ^{4} } -\eta ^{2} \right)+\frac{4\omega _{3}}{ \gamma ^{3} }, \label{thetaDot}
\end{eqnarray}
\end{subequations}
where  $m^{*}$ is a dimensionless polariton mass.

Phase diagrams for various dynamical regimes are determined by the property of polariton mass $m^{*} $ and by the sign of  Hamiltonian $H$ that is a conserved quantity.
At $m^{*} > 0,$ a polariton wave-packet exhibits diffusive and self-trapping regimes for which $\gamma \to \infty $, $\eta \to 0$ and $\gamma \to \text{constant}$ in the limit of infinite time scales ($t \to \infty $), respectively.

For an untitled trap of polaritons in the lattice, the momentum $p\left( t \right)=p_{0}$ is conserved.
By introducing the dimensionless polariton mass $m^{*} = \left( \frac{\partial^2 H}{\partial p^2} \right)^{-1}$, one can have  the effective Hamiltonian function $H$  in the dimensionless coordinates,
\begin{equation}
\label{effHam_norm}
H = -\omega _{1} \cos \left(p\right)e^{-\sigma } -\omega _{2} \cos \left(2p\right)e^{-4\sigma } + \frac{\omega _{3} }{\gamma },
\end{equation}
where $\sigma = \gamma ^{2} \eta ^{2}/8 +1/2 \gamma ^{2}$. The transition between different regimes is governed by an equation ${H\equiv H_{0} =0}$ that implies
\begin{equation}
\label{cos_of_ham}
\cos \left(p_{1,2}^{H_{0} } \right)\simeq -2\varepsilon _{10} \pm \sqrt{4\varepsilon _{10}^{2} +0.5},
\end{equation}
where $\varepsilon _{10} = e ^{3 \sigma _{0}} \Omega _{12} /8$; $\sigma _{0} \equiv \sigma (t=0)=1/2 \gamma ^{2} _0$ (we suppose that initially at $t=0$ $\gamma =\gamma _{0}$ and $\eta = 0$), $\Omega _{12} \equiv {\omega _{1} \mathord{\left/ {\vphantom {\omega _{1} \omega _{2} }} \right. \kern-\nulldelimiterspace} \omega _{2} } $, and we denote $H_{0} $ as initial value of the Hamiltonian $H$ that is, obviously, conserved quantity in the absence of dissipation.

Both of the roots \eqref{cos_of_ham} are located within the domain $-1\le \cos \left(p_{0} \right)\le 1$ if conditions
$
\gamma _{0} \ge \left( \frac{2}{3}{ \ln \left[\left|\Omega _{12} \right|^{-1} \right]} \right) ^{-1/2}$ and $ \left|\Omega _{12} \right| \le 1$ are fulfilled simultaneously.
Otherwise Eqs.~(\ref{cos_of_ham}) impose only one root.
It occurs for the tunneling rates $\left| \Omega _{12} \right| > 1$. Practically, this situation corresponds to large enough cavity sizes for which both of the tunneling rates are positive and $\omega _{2} $ vanishes rapidly.

Physically important bound state for our problem occurs in the domain of negative polariton mass and can be associated with soliton formation for the polariton wave-packet.
The polariton (bright) soliton wave-packet propagates with initial width ${\gamma _{0} }$, mass ${m^{*} =m_{0}^{*} <0}$ and velocity ${v_{g} = -\frac{\tan \left(p_{0} \right)}{\left|m_{0}^{*} \right|} + 4\omega _{2} \frac{\sin ^{3} \left(p_{0} \right)}{\cos \left(p_{0} \right)} e^{-4\sigma _{0} } }$ unchanged in time.
The mass of the soliton wave-packet can be found from
\begin{equation}
\label{eff_mass_norm}
\frac{1}{m_{0}^{*} } =\omega _{1} \cos \left(p_{0} \right)e^{-\sigma _{0} } +4\omega _{2} \cos \left(2p_{0} \right)e^{-4\sigma _{0} }.
\end{equation}
Strictly speaking, Eq.~\eqref{eff_mass_norm} defines characteristic domain
\begin{equation}
\label{cos_of_mass}
\cos \left(p_{1,2} \right)=0.5\left[-\varepsilon _{10} \pm \sqrt{2+\varepsilon _{10}^{2} } \right]
\end{equation}
of allowed wave-packet momentum where solitonic regime can be achieved.
It can be obtained under the conditions~$
\gamma _{0} \ge \left(\frac{2}{3}{\ln \left[4\left|\Omega _{12} \right|^{-1} \right]} \right) ^{-1/2}$ and $ \left|\Omega _{12} \right| \le 4$.

Solitons exist within the region for which inequalities $\cos \left(p_{2} \right)<\cos \left(p_{0} \right)<\cos \left(p_{1} \right)$ are hold for the positive tunneling rates $\omega _{1,2} $ ($\omega _{1,2} >0$) and for $\omega _{1} <0$ ($\varepsilon _{10} <0$), $\omega _{2} >0$.
Contrary, at $\omega _{1,2} <0$ solitons can be obtained at the outside of the named region.

In the limit of tight binding approximation ($\omega _{2} =0$) Eq. \eqref{cos_of_ham} implies only one solution $\cos\left(p_{1} \right)=0$ that corresponds to the physical situation described in~\cite{TrombettoniSmerziPRL2001,WangZhangZhangMaXuePRA810336072010} for atomic BEC lattice solitons.
In this case polariton solitons exist only for wave-packet momentum that obeys to inequality $\cos\left(p_{0} \right)<0$.

\section{\label{polWPdyn} Polariton wave-packet dynamics}

\begin{figure}
\centering
\includegraphics[width=8cm]{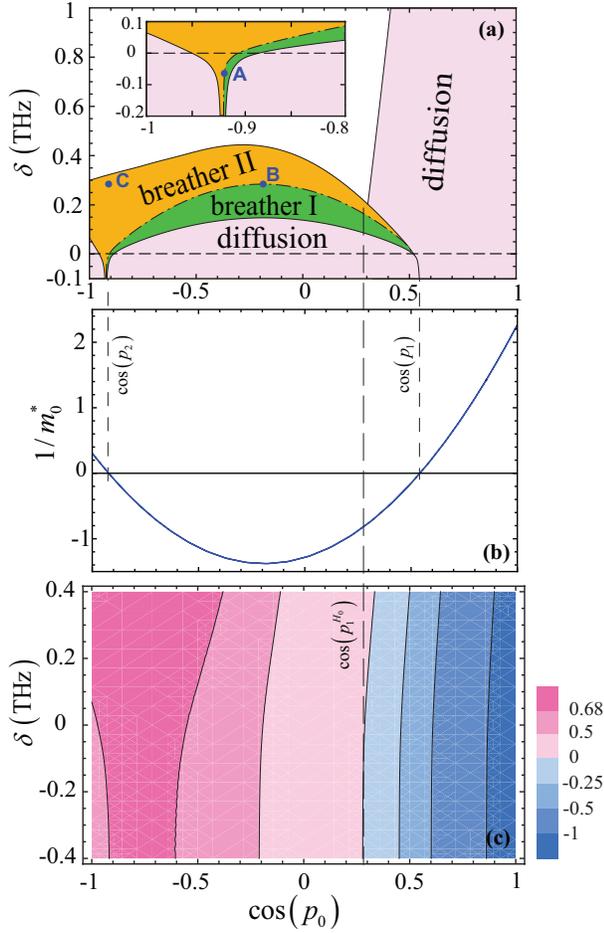}
\caption{\label{FIG_diagr_main} (Color online) (a) Dynamical phase diagram, (b) effective  polariton mass, shown in the inverse form ${1 \left/   \right.  m_{0}^{*} } $, and (c) the corresponding  Hamiltonian energy contour are shown in terms of the momentum $\cos \left(p_{0} \right)$ and detuning frequency $\delta$, respectively,  for the parameters $d=2 {\rm \mu m}$, $\gamma _{0} =5$, and $\eta _{0} =0$. The markers \textbf{A} and \textbf{B} (\textbf{C}) shown in (a) correspond to the polariton soliton (breather) states, which are used below for the storage and retrieval of optical information shown in Fig.~\ref{FIG_storage}.}
\end{figure}

It is much better to provide the analysis of polariton wave-packet dynamics in the dynamical phase diagram picture, that reflects particular features of polaritons in the lattice.
In Fig.~\ref{FIG_diagr_main}, we represented the corresponding dynamical phase diagram, the related polariton effective mass, and the Hamiltonian energy contours,  as functions of the momentum parameter $\cos \left(p_{0} \right)$.
For ${\cos \left(p_{0} \right)>\cos \left(p_{1} \right)}$, the initial polariton mass is positive and one can expect self-trapping and diffusive regimes only. The most important results can be obtained for LB polariton dynamics in other domains of momentum $p_{0}$.

\begin{figure}
\centering
\includegraphics[width=8cm]{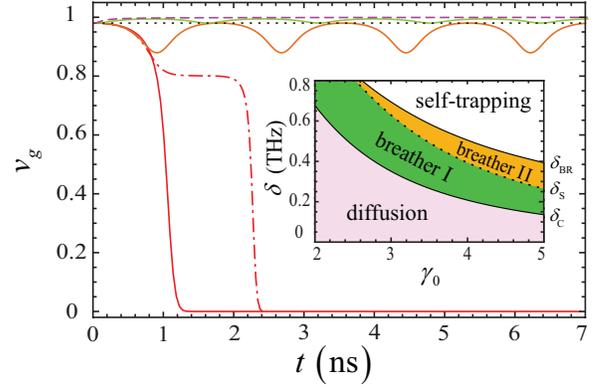}
\caption{\label{FIG_velos_Pi_div_BY_2} (Color online) The group velocity $v_{g} $ versus time $t$ for $\gamma _{0} =5$, $p_{0} =\pi /2$. Beginning from the top of the figure, $\delta  \equiv \delta _{\rm C} \approx 137.86 {\rm GHz}$ and $v_{0} \equiv v_{g} \left(t=0\right)=330214 {\rm m/s}$ (dashed curve); $210 {\rm GHz}$ and $v_{0} =148415 {\rm m/s}$ (green curve); ${\delta  =\delta _{\rm S}  \approx
%264.96
265 {\rm GHz}}$ and ${v_{0} =94379 {\rm m/s}}$ (dotted line); $380 {\rm GHz}$ and $v_{0} =46382 {\rm m/s}$ (brown curve); $\delta =\delta _{\rm BR} \approx 393.66 {\rm GHz}$ and $v_{0} =43250 {\rm m/s}$ (dashed-dotted red curve); $500 {\rm GH}z$ and $v_{0} =26907 {\rm m/s}$ (solid red curve). In the inset dependence of $\delta$ versus $\gamma _{0}$ for $p_{0}=\pi /2$ is plotted. }
\end{figure}

In the region $\cos \left(p_{2} \right)<\cos \left(p_{0} \right)<\cos \left(p_{1}^{H_{0} } \right)$ we have $m_{0}^{*} <0$, $H_{0} >0$ for all values of detuning $\delta $ -- see~Fig.~\ref{FIG_diagr_main}. Figure~\ref{FIG_velos_Pi_div_BY_2} demonstrates typical temporal dynamics of the wave-packet group velocity $v_{g} $ in this case.
In the inset the dependence of detuning $\delta$ as a function of initial width $\gamma _{0}$ of the wave-packet is presented.
For $\delta <\delta _{\rm C} $ we deal with the diffusive regime for which the group velocity tends to the constant value ${v_{g} \approx \sin \left(p_{0} \right)+2\omega _{2} \sin \left(2p_{0} \right)}$ asymptotically. On the other hand the group velocity oscillates in time within the window $\delta _{\rm C} <\delta <\delta _{\rm BR} $. For $\delta >\delta _{\rm BR} $, i.e. for the self-trapping regime $v_{g} $ rapidly vanishes and goes to zero. The soliton regime occurs for atom-field detuning $\delta =\delta _{\rm S} $ and obviously is characterized by a constant value of the group velocity -- dotted line in Fig.~\ref{FIG_velos_Pi_div_BY_2}.

\begin{figure}
\centering
\includegraphics[width=8cm]{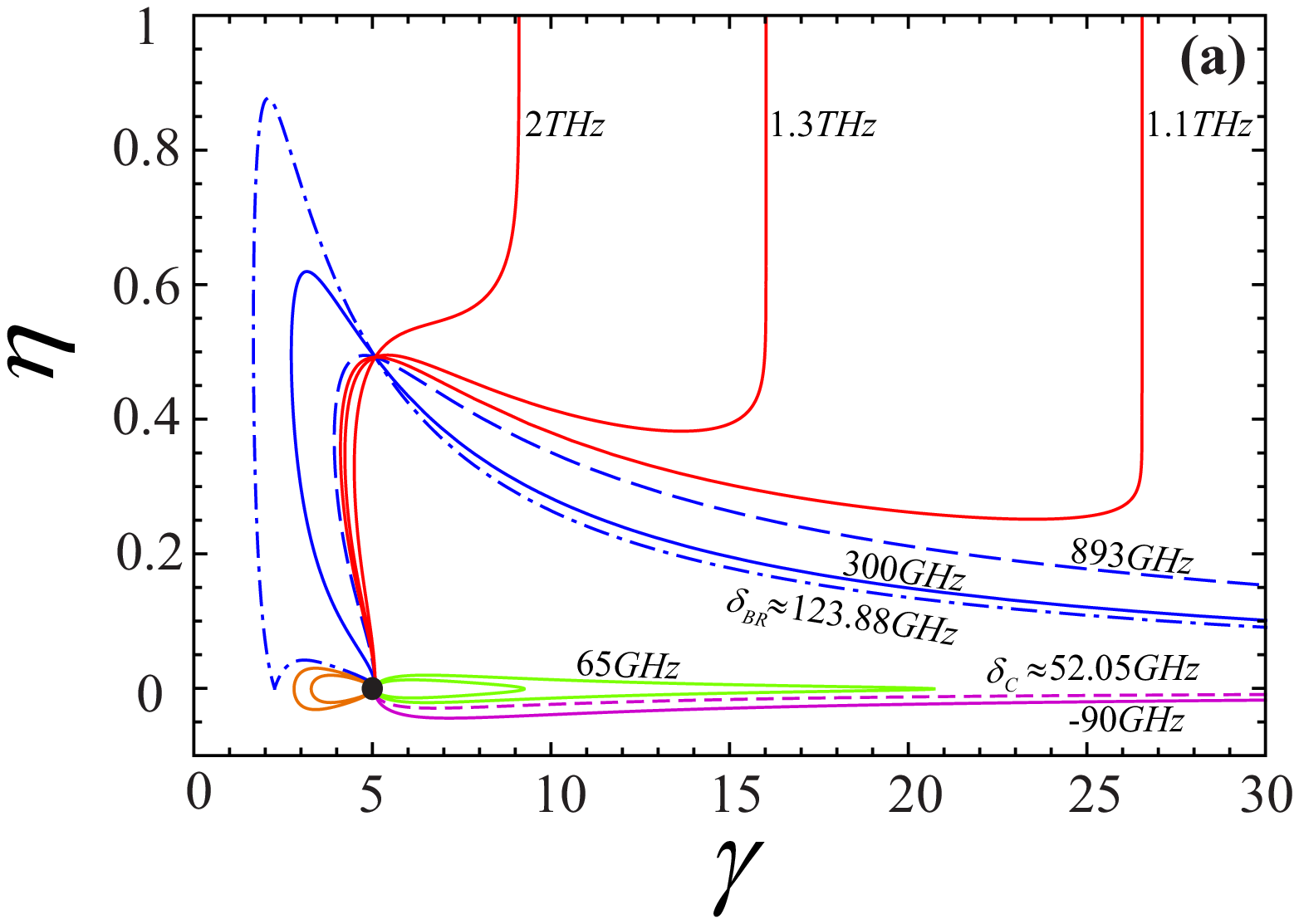}
\includegraphics[width=8cm]{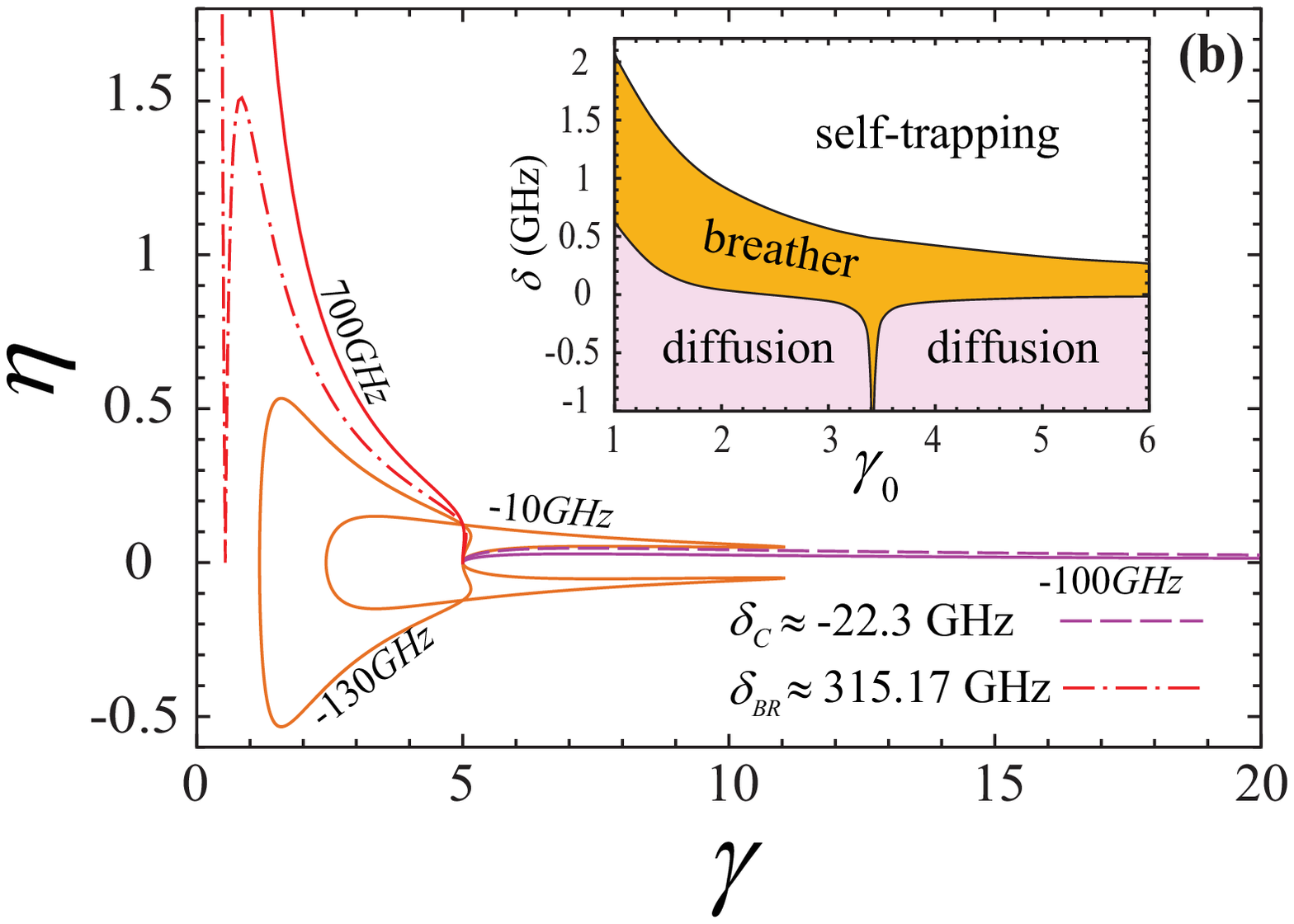}
\caption{\label{FIG_portr_cos_M0d94} (Color online) Trajectories in the $\eta -\gamma $ plane for various atom-field detuning $\delta $ with initial conditions $\gamma _{0} =5$, $\eta _{0} =0$ for (a) -- $\cos \left(p_{0} \right)=0.4$  and (b) --  $\cos \left(p_{0} \right)=-0.94$. The magnitudes of $\delta$ in (a) are $80 {\rm GHz}$ (for an unlabeled green curve), $115 {\rm GHz}$ and $120 {\rm GHz}$ (for inside and outside brown curves, respectively). Black dot in (a) corresponds to solitonic regime of the wave-packet parameters for $\delta = \delta _{\rm S}  \approx 99.24 {\rm GHz}$. }
\end{figure}

Essentially new results can be obtained when momentum obeys to the condition $\cos \left(p_{1}^{H_{0} } \right)<\cos \left(p_{0} \right)<\cos \left(p_{1} \right)$ and represents a narrow area in Fig.~\ref{FIG_diagr_main}.
Analysis of the polariton wave-packet dynamics in the discussed domain is straightforward. Due to the energy conservation law it is possible to establish a simple inequality $ \frac{\omega _{3} }{\gamma } -H_{\rm eff} >0$,
where $H_{\rm eff} \equiv H_{0} -\omega _{2} \left|\cos \left(2p_{0} \right)\right|e^{-4\sigma }$ can be recognized as a shifted Hamiltonian in this case.
The self-trapping regime for the wave-packet can be found out for $H_{\rm eff} > 0$ -- upper (red) curves in Fig.~\ref{FIG_portr_cos_M0d94}(a).
However, since $\sigma \gg 1$ we can suppose that $H_{\rm eff} \approx H_{0} $ in this limit. The maximal value $\gamma _{\max } \simeq {\omega _{3} \mathord{\left/ {\vphantom {\omega _{3} H_{0} }} \right. \kern-\nulldelimiterspace} H_{0} } $ of the width of the polariton wave-packet for the self-trapping regime can be obtained as a result. The set of other regimes are achieved at $H_{\rm eff} < 0$, or simply, at $H_{0} <0$.
By using the Hamiltonian $H_{0} $ we can arrive at an equation $\eta ^{2} =\frac{8}{\gamma ^{2} } \ln \left[\frac{\cos \left(p_{0} \right)}{\omega _{3} /\gamma -\left|H_{0} \right|} \right]$ for the curvature parameter $\eta $ that describes wave-packet behavior at its large width for $\gamma \gg 1$.
Since $m_{0}^{*} <0$ the system supports bright polariton soliton solutions and breather regimes as well. In particular, from the energy conservation law we can establish a relation $\frac{\omega _{3} }{\gamma } = H_{\rm eff} +\cos \left(p_{0} \right)e^{-\sigma } >0$. Hence, the lower diffusive regime  with ${\gamma \to \infty }$ and an equation ${\eta =\frac{2}{\gamma } \sqrt{2\log \left[{\cos (p_{0} )\mathord{\left/ {\vphantom {\cos (p_{0} ) \left|H_{\rm eff} \right|}} \right. \kern-\nulldelimiterspace} \left|H_{\rm eff} \right|} \right]} \to 0}$ occurs for ${\left|H_{\rm eff} \right|<\cos \left(p_{0} \right)}$.

On the other hand, if $\left|H_{\rm eff} \right|>\cos \left(p_{0} \right)$, the width $\gamma $ has to remain finite that corresponds to breather regimes. Transition between two regimes can be found out from an equation
$
\omega _{\rm 3,C} =\gamma _{0} \left[\omega _{2} \left|\cos \left(2p_{0} \right)\right|\left(1-e^{-4\sigma _{0} } \right) - \cos \left(p_{0} \right)\left(1-e^{-\sigma _{0} } \right)\right]
$.
It implies the critical number of polaritons that is characterized by the critical two-body polariton-polariton scattering parameter $\omega _{\rm 3,C} $ or relevant atom-field detuning $\delta _{\rm C} $.

Another physically interesting region of the wave-packet dynamics is characterized by the momentum domain ${\cos \left(p_{0} \right)<\cos \left(p_{2} \right)}$ that corresponds to a picture in the inset of Fig.~\ref{FIG_diagr_main}. Figure~\ref{FIG_portr_cos_M0d94}b demonstrates trajectories in the $\eta -\gamma $ space for the wave-packet parameters.

In this limit we deal with initially positive polariton mass ($m_{0}^{*} >0$) and the Hamiltonian ${H_{0} >0}$ -- see Fig.~\ref{FIG_diagr_main} for any values of detuning $\delta $. Proceeding as for previous cases we can find out a critical value of the two-body polariton scattering nonlinear parameter $\omega _{3,\rm C} =\gamma _{0} \left[\left|\cos \left(p_{0} \right)\right|\left(1-e^{-\sigma _{0} } \right)-\omega _{2} \cos \left(2p_{0} \right)\left(1-e^{-4\sigma _{0} } \right)\right]$ that separates polariton diffusive and localized regimes from each other. The polariton wave-packet being at the diffusive regime demonstrates approximately constant group velocity $v_{g} \approx \sin \left(p_{0} \right)+2\omega _{2} \left|\sin \left(2p_{0} \right)\right|$ at large times. It is important to note significantly nonlinear behavior of the polariton wave-packet parameters at the breathing region being under discussion. The polartion wave-packet width oscillates between the values $\gamma _{\min } $ and $\gamma _{\max } $ which are \textit{independent} on initial value $\gamma _{0}$. The group velocity also undergoes large amplitude nonlinear oscillations. However, the amplitude of oscillations is suppressed if we move toward the self-trapping area (cf.~Fig.~\ref{FIG_velos_Pi_div_BY_2}). In the limiting case for large enough detuning $\delta $ the atom-like LB polariton packet becomes self-trapped and ``stopped'' imposing vanishing group velocity $v_{g} \approx \sin \left(p_{0} \right)e^{-\gamma ^{2} \eta ^{2} /8} +2\omega _{2} \left|\sin \left(2p_{0} \right)\right|e^{-\gamma ^{2} \eta ^{2} /2} \to 0$.

\section{\label{Quantumstorageofopticalinformation}Tunneling-assisted optical information storage}
\begin{figure}
\centering
\includegraphics[width=8cm]{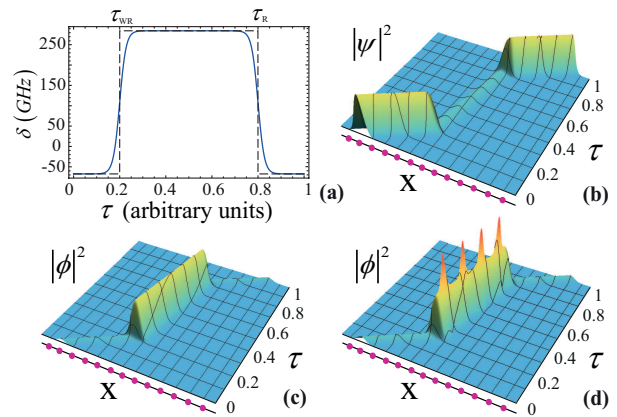}
\caption{\label{FIG_storage} (Color online) (a) Manipulation for the wave-packet is demonstrated through a time-dependent frequency detuning $\delta (t)$, for the lattice polariton solition solutions marked in Fig.~\ref{FIG_diagr_main}. Parameter $\chi=2 \pi \times 500 {\rm MHz}$.  Spatial-temporal evolution in a wave-packet storage and retrieval for the lattice polariton soliton, with the components shown schematically for the photon-like part $\left|\psi \right|^{2}$ in (b), and atomic excitation part $\left|\phi \right|^{2} $ for soliton (c) and breather (d), respectively. Here, the cavities  are shown as the purple points in $X$ direction. }
\end{figure}

The dynamical phase diagram for supporting different states permits physical protocol for adiabatic optical information storage and retrieval with the help of photon-like and matter-like duality of lattice polariton wave-packets; the protocol basing on the so-called rapid adiabatic passage (RAP) approach, which is slow on the time scale $(2g)^{-1}$ and fast enough in comparison with any incoherent process occuring in the atom-light system~\cite{MalinovskyKrauseEurPhysD142001}. In particular, atom-light detuning $\delta$ is a vital (governed) parameter in this case,~cf.~\cite{Leksin2007,HeshamiGreenYangHanPRA862012}.

The wave-packet for LB polaritons can be represented as
$\Psi \equiv \langle \Xi _2 \rangle = X \phi -C \psi$, with the wave functions $\phi$ and $\psi$ of the atomic exitation and optical field, respectively~--~see~\eqref{Xi_1_Xi_2}.
In particular, for a positive and large frequency detuning, $\delta \equiv \delta _{\rm at}$,  one  has a ``slow'' (matter-like) polariton solution, $\Psi \approx \phi$ in the cavity array; while for a negative frequency detuning,  $\delta \equiv \delta _{\rm ph}$, one has a ``fast'' (photon-like) soliton, $\Psi \approx \psi$.
In general, for the adiabatic storage of optical information we choose time dependent detuning such  as $\delta \left( t \right) = \delta _{\rm ph} +\frac{\delta _{\rm at} - \delta _{\rm ph}}{2} \left\lbrace  \tanh \left[ \chi \left( t - \tau_{ \rm WR} \right] \right)  - \tanh \left[ \chi \left( t - \tau _{\rm R}\right)  \right]   \right\rbrace  $, where parameter $\chi$ characterises the rate of detuning $\delta \left( t \right) $  variation; $ \tau _{\rm WR}$ ( $\tau _{\rm R}$) is a writting (retrieving) time moment.
In our problem the RAP approach requires the fulfillment of the condition~(cf.~\eqref{strongcouplingIneq})
\begin{equation}
\label{GGchi2g}
\max \left\lbrace \Gamma _{\rm at}, \Gamma _{\rm ph} \right\rbrace < \chi  < 2g.
\end{equation}

At the same time, one can require the fulfillment of the adiabaticity condition represented in the form~\cite{MalinovskyKrauseEurPhysD142001}
\begin{equation}
\label{adiCond}
\frac{4 \pi g \left|\dot{\delta }\right|}{\left[\left(2\pi \delta \right)^{2} +4g^{2} \right]^{3/2} } \ll 1
\end{equation}
and formulated for a two-level system that interacts with the external field.
It is important to note that, for the storage protocol with rubidium atoms in the cavity arrays, the conditions required in Eq.\eqref{GGchi2g} and Eq. \eqref{adiCond} are satisfied simultaneously at the rates $\chi < 2 \pi \times
%19.75
20 {\rm GHz}$.

As an example, here we establish two possibilities for the optical information storage.

First, consider the supported soliton that is a steady-state solution indicated by the markers \textbf{A} and \textbf{B} in Fig.~\ref{FIG_diagr_main}(a), both of which are bounded by the breather states, but with different wave-packet momenta.
At the writing stage,  such a wave-packet in the form of a polariton soliton  enters the configuration of the cavity array completely (writing time $\tau _{\rm WR}$ is about $1 {\rm ns}$ for Fig.~\ref{FIG_storage}(a), which operates with the  initial width of a polariton wave-packet equal to $10 {\rm \mu m}$); the polariton being {\it photon-like} and having the momentum $p_{0}^{\rm ph} = - \arccos \left( -0.922 \right)$ and the detuning frequency $\delta \equiv \delta _{\rm ph} \approx
%-67.49
-67.5 {\rm GHz}$, see Fig.~\ref{FIG_storage}(b).
Then, by adiabatical switching of the matter-light detuning frequency to the magnitude $\delta \equiv \delta _{\rm at} \approx   284.13 {\rm GHz}$,  {\it i.e.}, the corresponding lattice soliton solution moves across the phase boundary toward the marker \textbf{B}, resulting in possessing a low enough group velocity.
In this way, the original photon-like lattice polariton soliton is transferred into a {\it matter-like} one, with the momentum  $p_{0}^{\rm at} =\arccos \left(-0.2\right)$.
The  mapping of optical information from an incident optical field into coherent matter excitation is demonstrated in Fig.~\ref{FIG_storage}(c).
By reversing the detuning frequency adiabatically, the original wave-packet can be reconstructed back to the photon-like polariton soliton at the output of the cavity array.

Second, the possibility to arrange optical information storage involves the mapping of photon-like polariton soliton onto the  dynamically localized wave-packet state that is a breather state in our case.
In~Fig.~\ref{FIG_storage}(d) we demonstrate the mapping of photon-like solitonic polariton wave-packets into atomic excitations representing atom-like breather polariton wave-packet and characterized by point \textbf{C} in~Fig.~\ref{FIG_diagr_main}(a).
Maximally accessible (positive) vaue of detuning $\delta$ in this case is determined by the boundary value between two dynamical regimes that is self-trapping and breather II states in Fig.~\ref{FIG_diagr_main}(a)
The main advantage of the usage of breather polariton wave-packets for optical storage purposes is connected with the fact that at all storage stages we consider a polariton wave-packet with the same momentum, that is  $p _{0}=-\arccos(-0.922)$ for point \textbf{C}.
At the same time practical difficulties lie in the fact that it is necessary to select the appropriate retrieving time $\tau _{\rm R}$ according to the cycle of atomic breather evolution  for mapping back to photon-like polariton soliton.

\begin{figure}
\centering
\includegraphics[width=8cm]{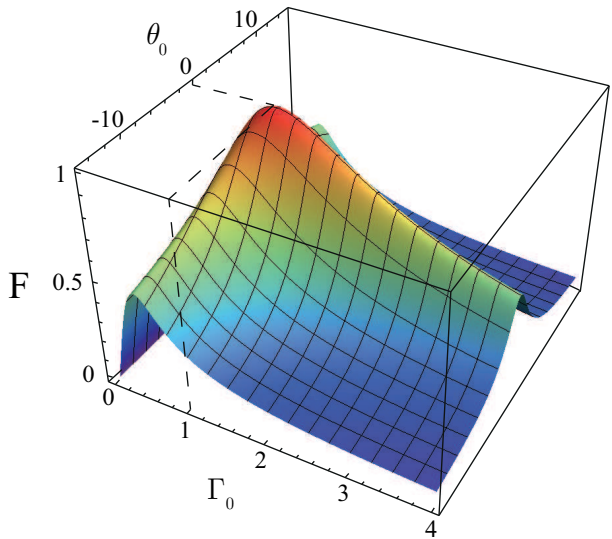}
\caption{\label{FIG_fidelonparameters} (Color online) Fidelity $F$ versus relative wave-packet width $\Gamma _0 = \Gamma _{\rm out} / \Gamma _{\rm in}$ and parameter $\theta _0 = \Gamma  ^{2} _{\rm in} \left( \theta _{\rm out} - \theta _{\rm in} \right)$.}
\end{figure}

Let us briefly discuss fidelity criteria that determines optimal polaritonic dynamical regimes for writing process.
It implies that we  take polariton wave-packet at different times being the closest to $\tau _{\rm WR} $, see~Fig.~\ref{FIG_storage}(a).
In our case, \textit{i. e.}, a pure quantum state fidelity can be simply recognized as the overlapping of states before  ($\Psi _{\rm \text{in}}$) and  after  ($\Psi _{\rm \text{out}}$)   writing~\cite{JoszaJModOpt411994,ScutaruJPhysAMathGen311998}:
\begin{equation}
\label{fidel_def}
F=\left|\int \Psi _{\rm in}^{*}  \Psi _{\rm out} d x \right|^{2},
\end{equation}
where  wave functions $\Psi _{\rm in , out}$ are established  in the $x$-space domain as \begin{equation}
\label{PsiStationary}
\Psi _{\rm in, out} =\left(\frac{2}{\pi \Gamma _{\rm in, out}^{2} } \right)^{1/4} \exp \left[-\left(\frac{1}{\Gamma _{\rm in, out}^{2} } -i\frac{\theta _{\rm in, out} }{2} \right)x^{2} \right],
\end{equation}
where $x=nd$,  $\Gamma =d\gamma $,  and $\theta =\eta /d^{2} $.
Note that  in the definition Eq.\eqref{PsiStationary} we have ignored the initial coordinate of a polariton wave-packet that is unimportant in this case.
Performing the integration in Eq.~\eqref{fidel_def},  we can get
\begin{equation}
\label{fidelCalculated}
F=\sqrt{\frac{16\Gamma _{0}^{2} }{4\left(1+\Gamma _{0}^{2} \right)^{2} +\Gamma _{0}^{4} \theta _{0}^{2} } },
\end{equation}
where $\Gamma _{0} =\Gamma _{\rm out} /\Gamma _{\rm in} $ and   $\theta _{0} =\Gamma _{\rm in}^{2} \left(\theta _{\rm out} -\theta _{\rm in} \right)$.

In Fig.~\ref{FIG_fidelonparameters}, the fidelity $F$ for various polariton dynamical regimes is examined.
The maximal value  $F=1$  is achieved for switching between two steady-state soliton regimes for the polariton wave-packet, with  $\theta _{0} =0$ and  $\Gamma _{0} =1$.
On the other hand,  $F$ vanishes  and goes to zero for the transitions involving self-trapping  ($\theta _{0} \to \pm \infty $) or diffusive ($ \Gamma _{0} \to \infty $) regimes.
Moreover, the local maxima in Fig.~\ref{FIG_fidelonparameters} obtained at $\theta _{0} = 0$ and $\Gamma _{0} \ne 1$ correspond to breather states of the polariton wave-packet, which can be used for dynamical optical information storage.

\section{\label{concls} Conclusions}
In summary, we consider the formation of lattice polariton solitons in the array of weakly coupled cavity-QED arrays, with the ensembles of two-level atoms embedded in each cavity.
With the introduction of the next-nearest photonic tunning effects, five different dynamical regimes are revealed; they including the diffusion, self-trapping, soliton, and two breather states.
Transformation between matter-like and photon-like  lattice polariton solitons paves the way to the  storage and retrieval of optical information through the  adiabatic manipulation of detuning frequency.
Obviously, the quantum optical information can be stored within the time interval that practically depends on qubit decoherence time and  Q-factor of cavity array.
However,  our protocol of the storage of optical information with the help of localized (soliton and/or breather)  states have some important advantages, compared to those with Gaussian-type optical pulses.
First of all, solitons are much more robust in respect of small perturbations, even  in the presence of small dissipation and decoherence effects~\cite{ChenLinLai2012,KarpmanSolPertrTheor}.
Second, if dissipation and decoherence effects become significant, then it will be possible to find some specific regimes for dissipation solitons supported.
In this limit, solitons are formed due to some additional optical pumping~\cite{PeschelEgorovLedererOptLet2004}.
Quantum properties of these solitons and the related robustness  in their quantum states against decoherence and dissipation effects would be very important~\cite{DantanCvilinskiPinardPRA2006}.
These problems should be considered separately and will be examined by us in the forthcoming  papers.

This work was supported by RFBR Grants No.~10-02-13300, No.~11-02-97513, No.~12-02-31601, No.~12-02-90419 and No.~12-02-97529 and by the Russian Ministry of Education and Science under Contracts 14.740.11.0700.

\end{document}